\documentclass[10pt]{iopart}

\usepackage{iopams}
\usepackage{amsgen}
\usepackage{amsfonts}
\usepackage{amsbsy}
\usepackage{amssymb}

\usepackage{tabularx}
\usepackage{graphicx}
\usepackage{times}
\usepackage{amssymb}
\usepackage{textcomp}
\usepackage{xcolor}
\usepackage[T1]{fontenc}
\usepackage{ulem}

\begin{document}
\title{Effective second-order correlation function and single-photon detection}
\author{P. Gr\"unwald}
\address{Aarhus Universitet, Institut for Fysik og Astronomi, Ny Munkegade 120, 8000 Aarhus C, Denmark.}
\ead{Electronic address: peter.gruenwald@phys.au.dk}

\date{\today} 

\bibliographystyle{iopart-num}

\begin{abstract}
Quantum-optical research on semiconductor single-photon sources puts special emphasis on the measurement of the second-order correlation function $g^{(2)}(\tau)$, arguing that $g^{(2)}(0)<1/2$ implies the source field represents a good single-photon light source. 
We analyze the gain of information from $g^{(2)}(0)$ with respect to single photons. Any quantum state, for which the second-order correlation function falls below $1/2$, has a nonzero projection on the single-photon Fock state. The amplitude $p$ of this projection is arbitrary, independent of $g^{(2)}(0)$. However, one can extract a lower bound on the single-to-multi-photon-projection ratio. A vacuum contribution in the quantum state of light artificially increases the value of $g^{(2)}(0)$, cloaking actual single-photon projection. Thus, we propose an effective second-order correlation function $\tilde g^{(2)}(0)$, which takes the influence of vacuum into account and also yields lower and upper bounds on $p$. We consider the single-photon purity as a standard figure-of merit in experiments, reinterpret it within our results and provide an effective version of that physical quantity. Besides comparing different experimental and theoretical results, we also provide a possible measurement scheme for determining $\tilde g^{(2)}(0)$.
\end{abstract}

\submitto{\NJP}
\maketitle

\maketitle

\section{Introduction}
Single photons (SPs) are an essential tool in both quantum optics and quantum information.  This includes, besides many other things, device-independent quantum cryptography~\cite{Sangouard2012,Diamanti2016} or a photonic quantum network~\cite{Kimble2008,Lodahl2017}. In terms of the Hanbury-Brown Twiss (HBT) experiment, 
which measures the second-order correlation function
\begin{equation}
g^{(2)}(\tau)=\frac{\langle\hat a^\dagger(0)\hat a^\dagger(\tau)\hat a(\tau)\hat a(0)\rangle}{\langle\hat a^\dagger(0)\hat a(0)\rangle^2},
\end{equation}
with $\hat a(\hat a^\dagger)$ the annihilation(creation) operator of a single mode of the field, a deterministic SP source should have $g^{(2)}(0)=0$, as this quantity is connected to the probability of emitting two photons (or more) at the same time. The first SP light source was the fluorescence of single atoms, which provided an almost perfect SP source, but with low intensity and even lower collection efficiency. Rydberg atoms provide a better suitable source, where the light can be manipulated on the slow time scales of electronic decays. Nowadays they are implemented in many quantum-information protocols~\cite{Saffman}.

Other sources of SPs have been identified, such as spontaneous parametric down conversion, which already was used to generate pure SP states~\cite{Mosley2008}, quantum scissors based on applying teleportation to a coherent input~\cite{Babichev2003}, and strongly attenuated coherent states common in quantum cryptography~\cite{Huttner1995}. Recently, it was even proposed to generate SPs via quenching the vacuum~\cite{Burillo2019}. A large field of research is focused on quantum dots as artificial atoms, coupled to different micro- or nanostructured environments in solids~\cite{MichlerExp,MichlerBook,Shih-1,Shih-2}. These tailored many-body structures have been proposed as a powerfull tool for SP-based architectures.
Their controllable variety in optical properties can substantially widen the range of applicability of SPs~\cite{Vuckevic2012}. However, because of their complex inner structure, it is not straight forward to show the SP character of such light sources. 

In semiconductor experiments a light source is said to emit SPs if the second-order correlation function $g^{(2)}(\tau)$ fulfills $g^{(2)}(0)<1/2$~\cite{Michler2000}. This criterion is widely used in the semiconductor optics literature for the last 20 years~\cite{Vuckevic2012,Michler2000,Kaupp2016,Bogdanov2017,Giess2015,Schweickert2018,Santis16,Lodahl2016}. Throughout this time, the information gained from this quantity has varied a lot, ranging from single emitters~\cite{Michler2000,Kaupp2016,Bogdanov2017} to SPs, from a quantitative interpretation as in~\cite{Santis16} to a downplay of this quantification as in~\cite{Giess2015,Laussy16}. There is significant research in trying to lower $g^{(2)}(0)$ further towards ideal zero, see~\cite{Schweickert2018} and references therein. More recently, the single-photon purity $b=1-g^{(2)}(0)$~\cite{Lodahl2016,Schweickert2018} has become a frequent figure of merit. This quantity, however, is not without issues. A two-photon quantum state $|\psi\rangle=|2\rangle$ has $b=50\%$, but there is no projection on the SP Fock state. A coherent state $|\alpha\rangle$ with $\alpha\neq0$ on the other hand has $b=0$ but the SP-projection is never zero.

The impact of spectral filtering of the signal on the actual SP character of the light source was analyzed~\cite{Laussy16,Blay17}. This is particularly interesting for semiconductor light sources, where often pulsed excitation schemes are applied. The authors of these papers showed that due to filtering the value of $g^{(2)}(0)$ can be shifted, sensitively depending on the actual source of light. In particular, different ideal SP sources (two-level systems, driven in different ways) may yield varying non-zero values for $g^{(2)}(0)$. It seems necessary in this context to consider, how accurate $g^{(2)}(0)$ is actually determined in a pulsed setup in general, compare~\cite{Broadband15} for the general determination of quantum correlations from spectrally filtered fields in different setups. 
To avoid any confusion in notation, there are three seperate physical entities in this discussion. First, the quantum state of the emitter before emission, second the emitted fields, and third the detected fields, i.e. the detected quantum states of light. Assuming the detection has been performed properly in experiments, we obtain information of the Fock-state statistics of the quantum state of light that would be used in applications. In simple textbook systems, the statistics of this quantum state would be directly connected to the operators of the emitter~\cite{WelVo}. Yet, more complex emission processes like cascades make this connection indeterminable. Hence, we focus on the statistics of the quantum state of light, and call the projection of this state onto the single-photon Fock state single-photon projection (SPP).

From a quantum-optical viewpoint, the only statement connected to single photons and the second-order correlation function is the antibunching (and more generally sub-poisson photon statistics) condition $g^{(2)}(0)<1$. This criterion implies nonclassicality of the light and, roughly speaking, a more likely appearance of lower numbers of photons. Notwithstanding, also coherent states $g^{(2)}(0)=1$ always have a SPP. It should also be noted that $g^{(2)}(0)=0$ is only a necessary criterion for SPs, not a sufficient one.
Recently, a somewhat different interpretation of $g^{(2)}<1/2$ was given in~\cite{Zubizarreta2017}, where it was shown to imply that the average photon number of the quantum state is in that case limited by 2. This means that even states, which on average have more than one photon, may fall under this category. Moreover, using this additional information can be used to give bounds on different multi-photon projections.

This work has two major aims. On the one hand we give a thorough analysis of what information can be gained from a valid measurement of $g^{(2)}(0)$. We give a mathematical proof that $g^{(2)}(0)<1/2$ implies that there is a nonzero SPP in the quantum state of light. However, the absolute probability $p$ of obtaining one photon in a photon-number measurement remains arbitrary, independent of $g^{(2)}(0)$. Yet, we can derive an analytic lower bound for the ratio of single-to-multi-photon projection (SMPP). On the other hand, from our results we can conclude that vacuum contributions lead to an artificial enhancement of $g^{(2)}(0)$ cloaking SPPs. We use these results to propose a scaled effective second-order correlation function $\tilde g^{(2)}(0)$, which takes vacuum into account and substantially extends the range of applicability for SPP-detection. Moreover, as the vacuum contribution can be measured using click detectors, $\tilde g^{(2)}(0)$ can be determined in the lab. We give relative and absolute bounds on $p$, and analyze theoretical quantum states and real experiments on semiconductor SP sources. Furthermore, we reinterpret the SP purity based on these results. Finally, we analyze a possible setup to measure the effective second-order correlation function directly.
From now on, we will only focus on the value of $g^{(2)}(0)$ and omit the time argument in the notation.

The paper is organized as follows. In section~\ref{sec.Proof}, we give the full proof that $g^{(2)}<1/2$ indicates a nonzero projection on the SP Fock state in the analyzed quantum state of light. We then show in section~\ref{sec.AbsAmp} that the amplitude of this projection is arbitrary, independent of $g^{(2)}$. In section~\ref{sec.RelAmp} we derive a lower bound on the relative amplitude, which directly reveals the importance of the vacuum contribution and yields the effective second-order correlation function $\tilde g^{(2)}$ we propose. In section~\ref{sec.kS} we apply our results to known quantum states of light, showing the extended applicability of the original $g^{(2)}$ criterion. Some alternative interpretations to the above results are given in section~\ref{sec.alternative}. Section~\ref{sec.4} is devoted to comparing our results to previous work on specific semiconductor systems. We propose a simple measurement scheme based on post selection to obtain the effective correlation function in section~\ref{sec.5}. Finally, we give conclusions and an outlook in section~\ref{sec.out}.

\section{Proof that $g^{(2)}<1/2$ implies single-photon projection}\label{sec.Proof} 
Let us first give the full proof that when for a given state $\hat\varrho$ we obtain $g^{(2)}<1/2$, the state has a nonzero SPP $\langle 1|\hat\varrho|1\rangle$. For any given Fock state $|n\rangle$, $n\geq1$, one can easily calculate~\cite{WallsMilburn}
\begin{equation}
	g^{(2)}=\frac{n-1}{n}=1-\frac{1}{n}.
\end{equation}
Hence, for any Fock state with $n>1$, $g^{(2)}$ will be greater or equal to one half. For the vacuum state $|0\rangle$, one may use the limit of coherent states to define $g^{(2)}:=1$. Thus, all Fock states besides $|1\rangle$ follow the criterion $g^{(2)}\geq1/2$, while for $|1\rangle$, we have $g^{(2)}=0$. 

Any pure quantum state can be written in the Fock basis as a linear combination.
Consquently, if a superposition of two arbitrary states with disjoint Fock statistics can not yield a lower $g^{(2)}$ than either of its constituents, the proof of the above conjecture would be completed for pure states. Due to the diagonal correlations involved within $g^{(2)}$ superposing two pure quantum states with disjoint photon statistics 
\begin{equation*}
|\psi\rangle=\sqrt{s}|\psi_1\rangle+\sqrt{1-s}e^{i\phi}|\psi_2\rangle,\ s\in[0,1]
\end{equation*}
and the statistical, incoherent mixture of two arbitrary pure quantum states 
\begin{equation*}
\hat\varrho=s|\psi_1\rangle\langle\psi_1|+(1-s)|\psi_2\rangle\langle\psi_2|,
\end{equation*}
result in the same form for $g^{(2)}$,
\begin{equation}
	g^{(2)}=\frac{s\langle\psi_1|\hat a^{\dagger2}\hat a^2|\psi_1\rangle+(1-s)\langle\psi_2|\hat a^{\dagger2}\hat a^2|\psi_2\rangle}{[s\langle\psi_1|\hat a^{\dagger}\hat a|\psi_1\rangle+(1-s)\langle\psi_2|\hat a^{\dagger}\hat a|\psi_2\rangle]^2}.\label{eq.g2Super_1}
\end{equation}
Finally, if we do not use the fact that $|\psi_i\rangle$ are pure states and only consider the expectation values appearing in equation~(\ref{eq.g2Super_1}) themselves, we can substitute them with general density operators $\hat\varrho_i$. Hence, the proof of our conjecture would be done for all quantum states if we can show that the right-hand side of equation~(\ref{eq.g2Super_1}) can not become lower than the value of $g^{(2)}$ for either of the two states involved therein. Because of the nonlinear nature of $g^{(2)}$ with respect to the quantum state, this property has to be shown explicitely.

Defining $g_i=g^{(2)}$ of $\hat\varrho_i$ and $n_i=\textrm{Tr}\{\hat\varrho_i\hat a^\dagger\hat a\}$, $i=1,2$,
we can write equation~(\ref{eq.g2Super_1}) as
\begin{equation}
	g^{(2)}=\frac{sn_1^2g_1+(1-s)n_2^2g_2}{[sn_1+(1-s)n_2]^2}=\frac{sg_1+(1-s)r^2g_2}{[s+(1-s)r]^2},
\end{equation}
with $r=n_2/n_1\geq0$. Let us further assume without loss of generality that $g_2=tg_1$ with $t\in [0,1]$, i.e. $g_2\leq g_1$. The first derivative of $g^{(2)}$ with respect to $s$ reads as
\begin{eqnarray}
\frac{1}{g_1}\frac{dg^{(2)}}{ds}&=&\frac{d}{ds}\frac{s+(1-s)r^2t}{[s+(1-s)r]^2}\nonumber\\
&=&\frac{(1-r^2t)(s+(1-s)r)-2(1-r)(s+(1-s)r^2t)}{(s+(1-s)r)^3}\nonumber\\
&=&\frac{r[t(1-r)^2-t+1]-(1-r)(1-r^2t)s}{[s+(1-s)r]^3}.\label{eq.g2deriv}
\end{eqnarray}
The denominator of the derivative is positive for any value of $s$ and $r$. The numerator is linear in $s$. Thus, there can be no more than one extreme point. As $g^{(2)}$ can not decrease over a full shift of $s$ from 0 to 1, a value below $g_2$ thus requires a downward slope of $g^{(2)}$ at $s=0$. That slope can easily be seen from equation~(\ref{eq.g2deriv}) to be
\begin{equation}
\left.\frac{1}{g_1}\frac{dg^{(2)}}{ds}\right|_{s=0}=\frac{t(1-r)^2-t+1}{r^2}.
\end{equation}
The only combination for which this is not positive definite is $r=t=1$, In this case, the two constituent states have equal $g_i$ and $n_i$. Consequentially, they can not be distinguished by the incoherent superposition in $g^{(2)}$, and the second-order correlation function is simply constant. Hence, $g^{(2)}$ does not decrease at $s=0$ and no combination of states can lower the second-order correlation function below $g_2$. In other words, for $g^{(2)}<1/2$ we have a nonzero SPP, i.e. $\langle1|\hat\varrho|1\rangle>0$.

\begin{figure}[h]
\begin{center}
 	\includegraphics[width=8cm]{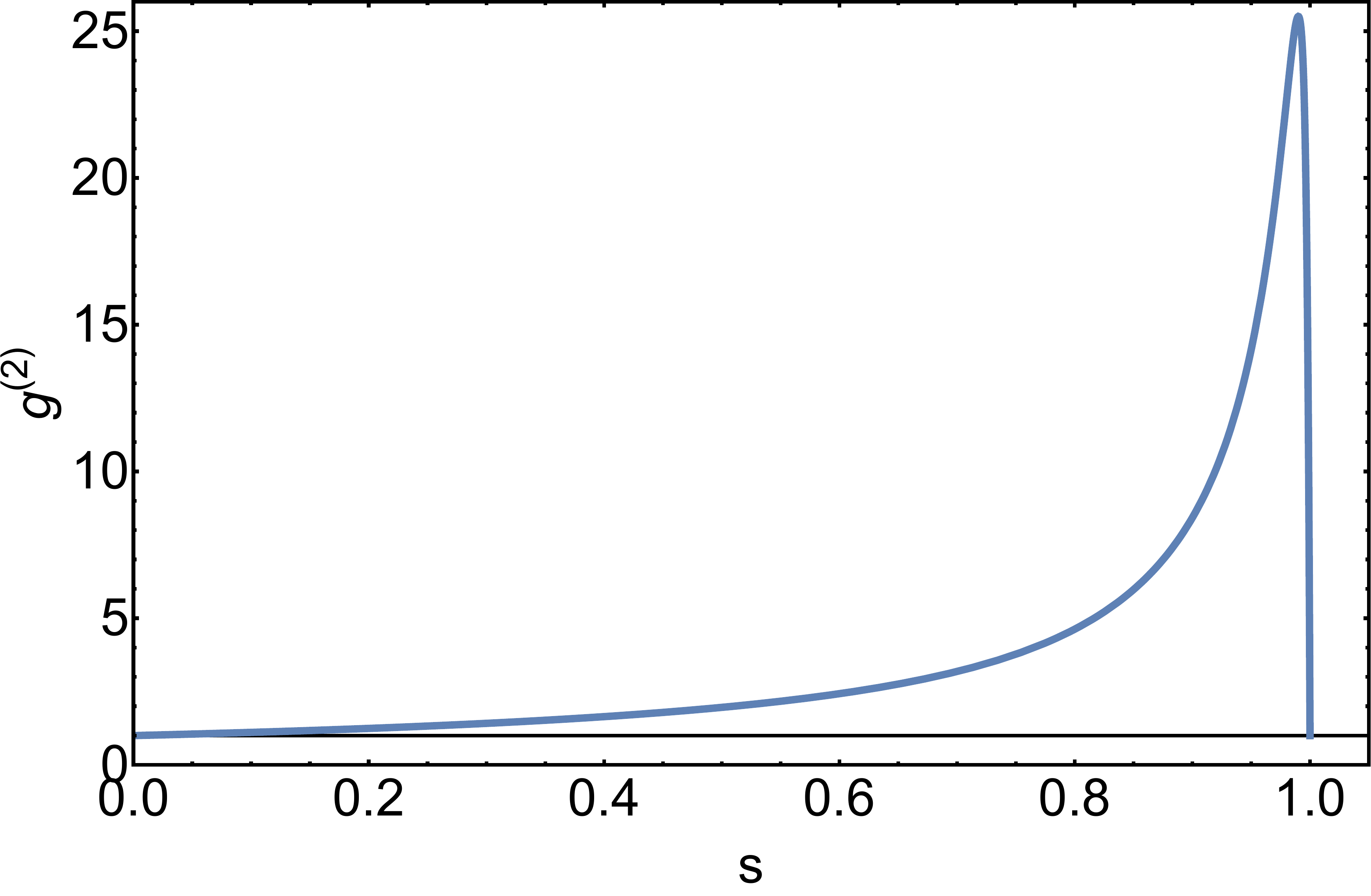}
	\end{center}
 	\caption{Overall second-order correlation function for an incoherent superposition with $g_1=g_2=1$ and $r=100$ over $s$.}
 	\label{fig.relg}
 \end{figure}

Note that the inverse is not true, and $g^{(2)}$ can become larger than either of its constituents. If we consider two states with $t=1$, that is, states with different average photon numbers but the same $g_i$, $g^{(2)}$ always increases above $g_i$ and maximizes for $s=r/(1+r)$. An interesting application of this case is given by two coherent states with different amplitudes. In this case we have $\hat\varrho_i=|\alpha_i\rangle\langle\alpha_i|$, $\alpha_i\in\mathbb C$ and the average photon numbers $n_i=|\alpha_i|^2$.
As both states have $g_i=1$, we can easily derive the value of $g^{(2)}$ for their incoherent superposition at the maximum to be
\begin{equation}
	g^{(2)}_\textrm{max}=\frac{(1+r)^2}{4r}\approx\frac{r}{4},\ r\gg1.
\end{equation}
That means, for two coherent states being statistically mixed - which represents a fully classical state - one with large, one with small coherent amplitude, the second-order correlation function scales up limitless. This is an example of a classical state with superbunching to arbitrary orders. 
In figure~\ref{fig.relg}, we depict $g^{(2)}$ for this case with $r=100$ over varying $p$. The same argument holds, e.g., for thermal states with $g_i=2$.

\section{Amplitude of single-photon projection}\label{sec.AbsAmp}
\begin{figure}[h]
\begin{center}
  \includegraphics[width=8cm]{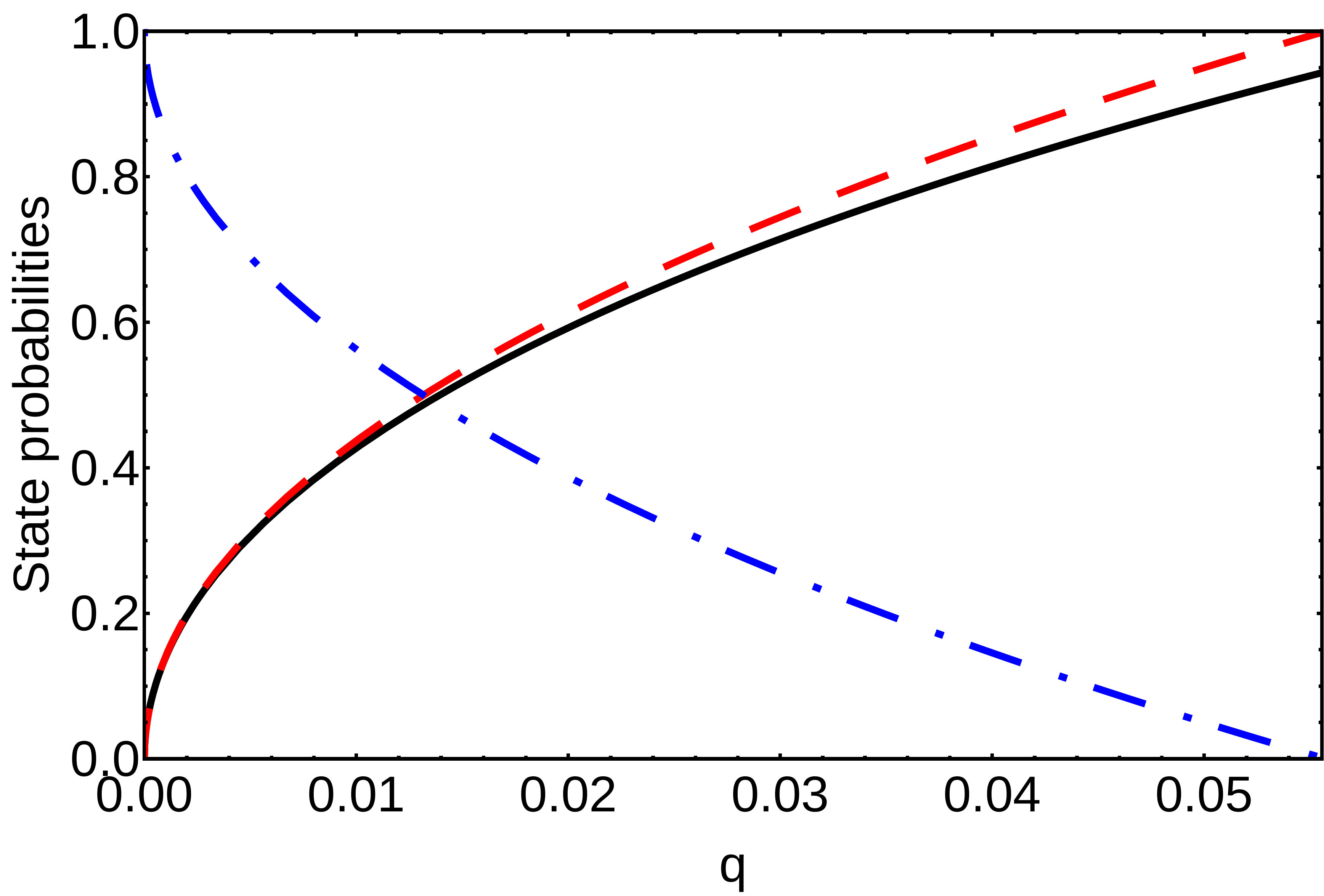}
\end{center}
  \caption{Different contributions $p$ (black, solid), $p+q$ (red, dashed), and vacuum $1-p-q$ (blue, dash-dotted) to the state in equation~(\ref{eq.psidef1}) for $g^{(2)}=0.1$.}\label{fig.exam_01}
\end{figure}

With the existence of a nonvanishing SPP shown, the next step is to quantify the amplitude $p$ of this projection. Unfortunately, including the effects of vacuum, it can be arbitrarily small, independent of $g^{(2)}$. Consider the following state:
\begin{equation}
	|\psi\rangle=\sqrt{1-p-q}|0\rangle+\sqrt{p}|1\rangle+\sqrt{q}|2\rangle.\label{eq.psidef1}
\end{equation}
Note that, due to the correlations under study being exclusively diagonal in Fock space, the phases in the prefactors are irrelevant. We easily compute
\begin{equation}
	g^{(2)}=\frac{2q}{(p+2q)^2}. \label{eq.g2_2}
\end{equation}
Solving equation~(\ref{eq.g2_2}) for $p$ yields
\begin{equation}
	p=\sqrt{\frac{2q}{g^{(2)}}}-2q.\label{eq.pq_1}
\end{equation}
The maximal value for $p$ is obviously $g^{(2)}$-dependent. However, $q$, and subsequently $p$ can be chosen arbitrarily small for any fixed $g^{(2)}$. For example, in figure~\ref{fig.exam_01}, we show $p$, $p+q$ and $1-p-q$ over $q$ for a fixed $g^{(2)}=0.1$. As one can see, there is no lower bound on $p$ other than zero.
Hence, for determining the absolute amplitude of the SPP, $g^{(2)}$ is insufficient, even as an approximation.

One can also show that $g^{(2)}$ gives no upper limit on $p$ either. Consider a state 
\begin{equation}
	|\tilde\psi\rangle=\sqrt{p}|1\rangle+\sqrt{1-p}|n\rangle\label{eq.psilargeg2}
\end{equation}
with fixed $g^{(2)}$ and variable Fock state number $n\geq2$. The second-order correlation function now reads
\begin{equation}
	g^{(2)}=\frac{n(n-1)(1-p)}{[p+n(1-p)]^2}.
\end{equation}
Inverting again for $p$ one obtains
\begin{equation}
	p=1-\frac{1}{(n-1)g^{(2)}}\left(\frac{n}{2}-g^{(2)}-\frac{n}{2}\sqrt{1-\frac{4}{n}g^{(2)}}\right).
\end{equation}
For sufficiently large $n$, $p$ gets arbitrarily close to 1, as
\begin{equation}
	p\approx1-\frac{g^{(2)}}{n(n-1)}.
\end{equation}
Note that even for $g^{(2)}>1/2$ we can have arbitrarily large $p$.
Thus, there is no information on the absolute probability to obtain SPs in the explicite value of $g^{(2)}$ itself. Note as well, if we fix $g^{(2)}=0.1$, for each of these states, the previously mentioned SP-purity would always be $b=1-g^{(2)}=90\%$, even though $p$ varies arbitrarily.

\section{Effective second-order correlation function}\label{sec.RelAmp}

\begin{figure}[h]
\begin{center}
  \includegraphics[width=8cm]{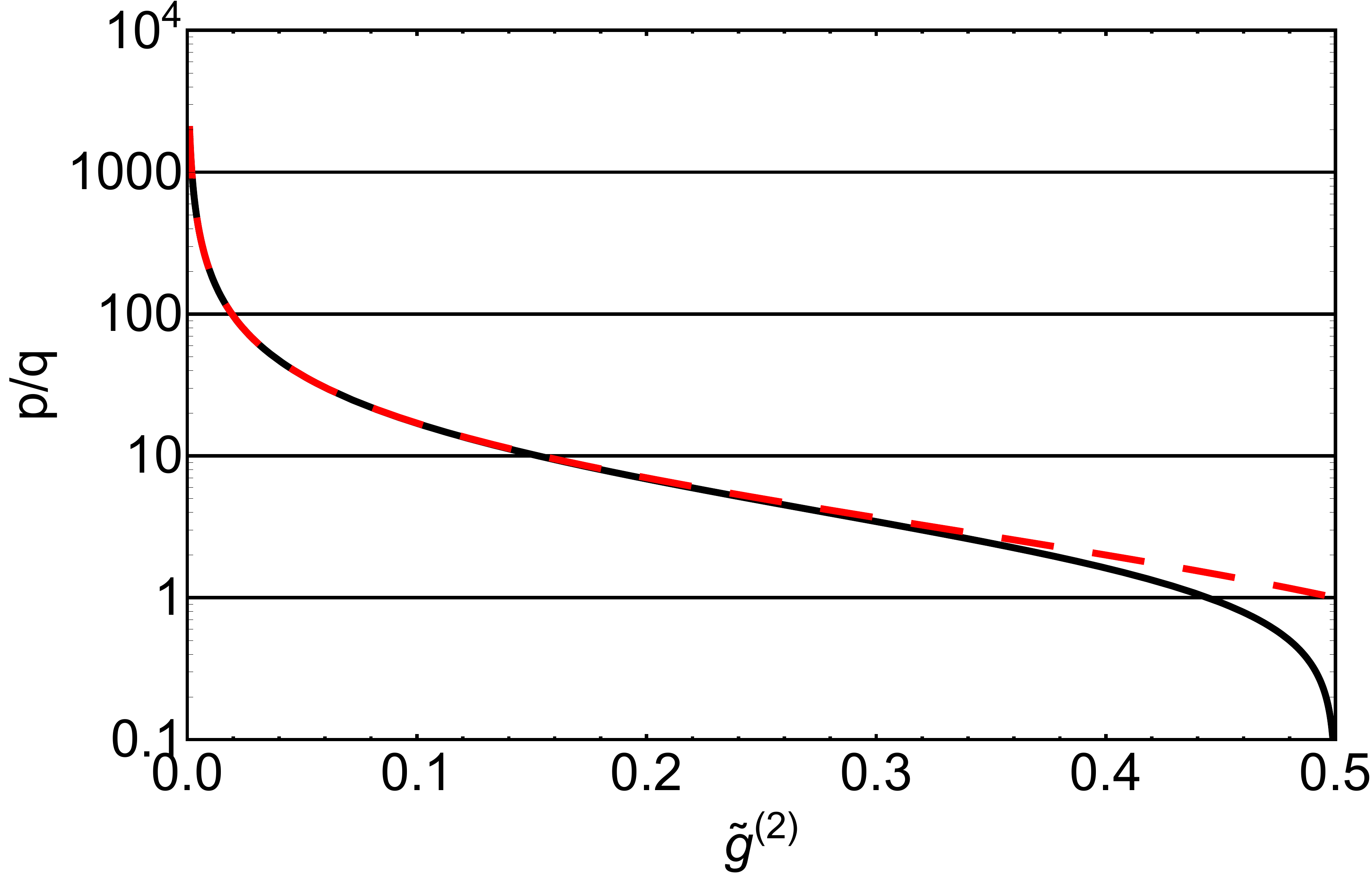}
\end{center}
  \caption{Lower bound for $p/q$, according to equation~(\ref{eq.lower_bound_g2}) (solid black line) as well as the approximation for $\tilde g^{(2)}\ll1$ (dashed red line).}\label{fig.lower_bound_1}
\end{figure}

Notwithstanding the above result, the value of $g^{(2)}$ does allow to give a lower bound for the ratio $p/q$, i.e. the ratio of single-to-multi-photon projection (SMPP). In most semiconductor scenarios it is desirable to have a source with high SMPP ratio. A similar bound was proposed in~\cite{Yamamoto2001} for pulsed excitations. Our result is a generalization based on the value $g^{(2)}$.
Consider now the state in equation~(\ref{eq.psidef1}) only with the state $|2\rangle$ being substituted for a general state $|\psi_2\rangle$ with
\begin{equation}
|\psi_2\rangle=\sum\limits_{k=2}^\infty c_k|k\rangle,
\end{equation}
covering all multi-photon projections of the state $|\psi\rangle$, which now reads as
\begin{equation}
	|\psi\rangle=\sqrt{1-p-q}|0\rangle+\sqrt{p}|1\rangle+\sqrt{q}|\psi_2\rangle.
\end{equation}
Again, this does not limit the generality of our results, as phases of orthogonal states are irrelevant for the diagonal correlations measured and general mixed states yield the same results. Analogous to equation~(\ref{eq.pq_1}), we find
\begin{eqnarray}
g^{(2)}&=&\frac{qn_2^2g_2}{(p+n_2q)^2}\\
\frac{p}{q}&=&n_2\left[\sqrt{\frac{g_2}{g^{(2)}q}}-1\right],
\end{eqnarray}
where $n_2$ and $g_2$ are the average photon number and second-order correlation function of $|\psi_2\rangle$. Applying the results of section~\ref{sec.Proof} onto $|\psi_2\rangle$ which has only multi-photon projections, we know $n_2\geq2$ and $g_2\geq1/2$, with the equality being given in both cases for $|\psi_2\rangle=|2\rangle$. Therefore, the ratio of $p/q$ has a lower bound of
\begin{equation}
\frac{p}{q}\geq2\left[\frac{1}{\sqrt{2g^{(2)}q}}-1\right].\label{eq.p/qapp}
\end{equation}
Note that already from $q\leq1$ on the right-hand side of equation~(\ref{eq.p/qapp}), we can deduce a nonzero lower bound for $p/q$ for $g^{(2)}<1/2$. Using $p=1-q-x$, $x$ being the vacuum projection, $x=|\langle 0|\psi\rangle|^2$ or in the general case $x=\langle0|\hat\varrho|0\rangle$, to solve for $q$ and reinserting the result on the right-hand side of equation~(\ref{eq.p/qapp}), we obtain the $q$-independent optimal value of this lower bound as
\begin{equation}
	\frac{p}{q}\geq 2\left[\frac{1}{\sqrt{2(1-\tilde g^{(2)}-\sqrt{1-2\tilde g^{(2)}})}}-1\right]=\frac{2\sqrt{1-2\tilde g^{(2)}}}{1-\sqrt{1-2\tilde g^{(2)}}}\label{eq.lower_bound_g2}
\end{equation}
with
\begin{equation}
	\tilde g^{(2)}=(1-x)g^{(2)}.
\end{equation}
The ratio is dependent on only one parameter, which is, however, not $g^{(2)}$ but a scaled version of it. We call $\tilde g^{(2)}$ the effective second-order correlation function. We find the following general statements: for $\tilde g^{(2)}=1/2$ this ratio becomes zero, as $p=0$ becomes possible. For $\tilde g^{(2)}\to0$ it becomes infinity as then $q\to0$. For $\tilde g^{(2)}\ll1$ the right-hand side of equation~(\ref{eq.lower_bound_g2}) can be approximated by $(2/\tilde g^{(2)})-3$. Both functions are depicted in figure~\ref{fig.lower_bound_1}.
The necessary $\tilde g^{(2)}$ to have a desired ratio $p/q\geq N$ is given by
\begin{equation}
	\tilde g^{(2)}\leq\frac{2(N+1)}{(N+2)^2}.
\end{equation}
For $\tilde g^{(2)}\geq4/9$ we may even have a larger multi-photon projection than SPP. Assuming a dominant SPP requires $N=10$, we find the upper limit of $\tilde g^{(2)}=11/72\approx0.15$.

\begin{figure}[h]
\begin{center}
  \includegraphics[width=8cm]{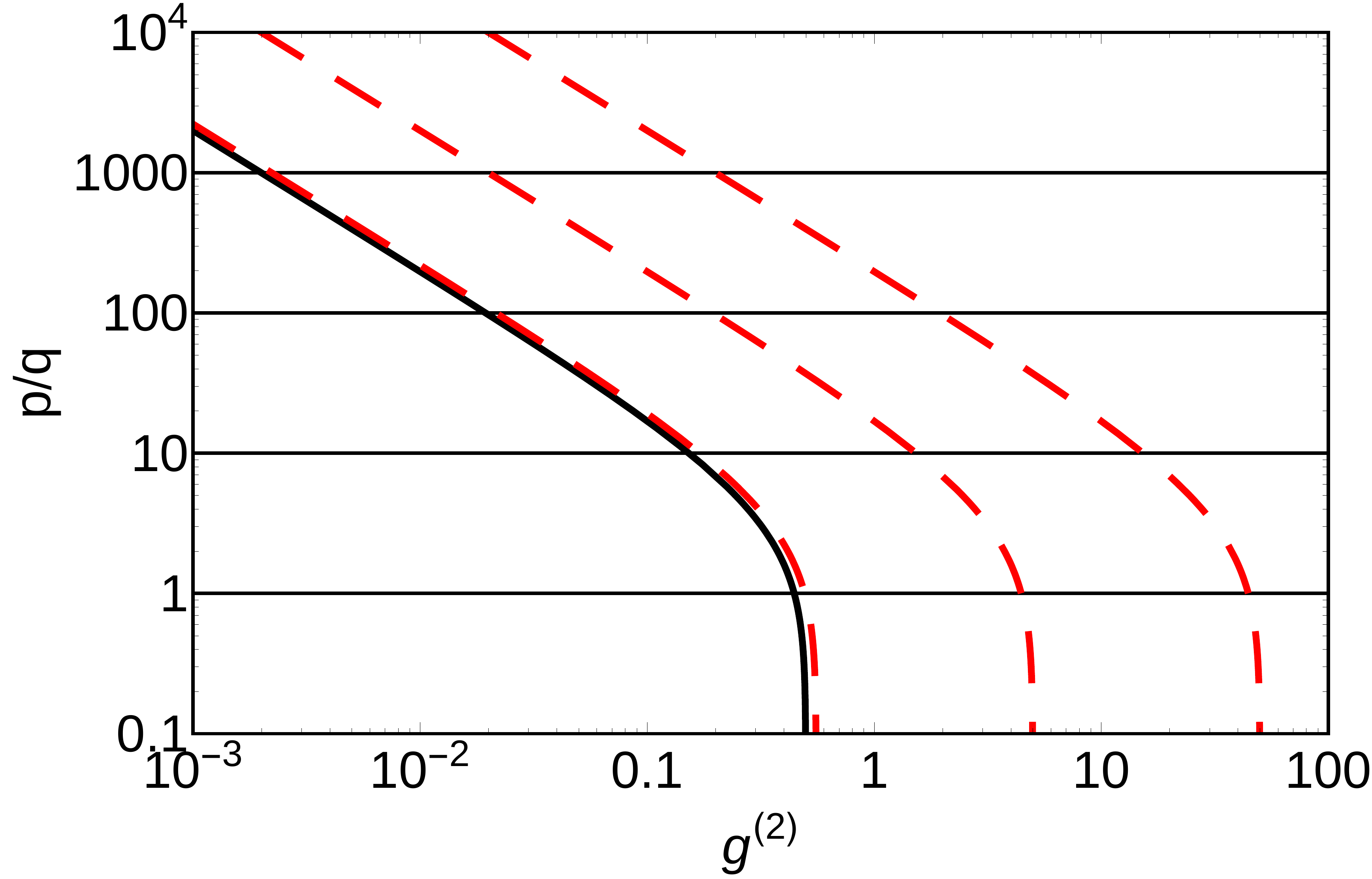}
\end{center}
  \caption{Lower bound for $p/q$, according to equation~(\ref{eq.lower_bound_g2}) for different $x$. From left to right: $x=0,0.1,0.9,0.99$.}\label{fig.lower_bound_2}
\end{figure}

Without knowledge of $x$ we have to assume $x=0$ and set $\tilde g^{(2)}=g^{(2)}$. This leads back to the previous statments for $g^{(2)}$. In other words, if in the experiments the vacuum contribution is not determined, equation~(\ref{eq.lower_bound_g2}) yields all that can be concluded from $g^{(2)}<1/2$. This does not exclude further conclusions like an upper bound on $q$ (see section~\ref{sec.alternative}) or on the average photon number~\cite{Zubizarreta2017}. It does provide one way to formulate this limited information quantitatively.
 If, however, a non-zero vacuum contribution is known, this knowledge can be extended, see figure~\ref{fig.lower_bound_2} for a visualization. For example, for a light source with 90\% vacuum ($x=0.9$), $g^{(2)}<5$ already yields nonzero SPP, while $g^{(2)}=0.5$ actually implies $p/q>37$, which could be considered a very good SP source with respect to the SMPP. On one hand, for a source with high $x$, one may now say that $g^{(2)}<1/2$ does imply to find predominantly SPs in photon-number experiments, rather than multiple photons. On the other hand, without definite information on the vacuum, the question of how good this source is, is undecidable. 
It should be clarified that large vacuum fluctuations counter the deterministic nature of a perfect SP source. If deterministic photons are needed for a given application, we can now clearly state
that additional information is required. The vacuum contribution $x$ is one example for such a quantity. For a nearly deterministic source, the conditions are $x\ll 1$, and $p/q\gg1$, simultaneously.
If instead only high SMPP rate is of interest, $g^{(2)}$ and $\tilde g^{(2)}$ provide versatile information.

There is a simple physical explanation for this huge influence of the vacuum contribution $x$ on the SMPP ratio of light.
Consider a state $\hat\varrho_0$ without vacuum ($x=0$) and a given SMPP ratio $p/q$. What happens to $g^{(2)}$, if we include vacuum $x$, but request that $p/q$ should remain fixed? Again, based on the diagonal correlations we use, we can include vacuum incoherently as
\begin{equation}
	\hat\varrho_x=x|0\rangle\langle0|+(1-x)\hat\varrho_0.\label{eq.vacincl}
\end{equation}
All ratios of single- and mutli-photon projections are fixed in this description.
Due to the linear combination of density matrix elements in arbitrary expectation values, $\langle\hat a^{\dagger n}\hat a^n\rangle$ is scaled down by a factor of $(1-x)$, if vacuum is included, independent of $n$. The second-order correlation function is a quotient of one expectation value with the square of another, leading to an overall upscaling of $g^{(2)}$ by a factor of $1/(1-x)$. This factor is exactly compensated in our effective second-order correlation function $\tilde g^{(2)}$ and thus yields the correct lower bound on $p/q$ independent of $x$. We rescale the correlation function to a vacuum-independent parameter.

Finally, the inclusion of the vacuum allows us to calculate absolute limits for the SPP $p$. Setting the right-hand side of equation~(\ref{eq.lower_bound_g2}) equal to $C$, we can use $q=1-x-p\geq0$ and obtain
\begin{equation}
	(1-x)\geq p\geq(1-x)\frac{C}{1+C}.\label{eq.abs_bounds}
\end{equation}	
Note that such an absolute boundary for $p$ was not possible without knowledge of the vacuum contribution as $x$ could be arbitrarily close to 1. Also note that in this notation, $\tilde g^{(2)}$ and $x$ enter the bounds independently. Thus, knowledge of the vacuum projection is crucial for determining $p$.

\section{Application to classical states}\label{sec.kS}	
Before we move on to apply our results to explicite previous works, it is insightful to look at known quantum states of light, to gauge the quality of these results. For this purpose we analyze our model for the classical cases of a coherent and a thermal state. Their density matrix is analytically known and with $g^{(2)}_\textrm{coh}=1$ and $g^{(2)}_\textrm{th}=2$, they would both violate the standard condition for nonzero SPP, $g^{(2)}\leq1/2$. Likewise, both states always have a SPP and for low average photon numbers $\langle\hat n\rangle\ll1$ approach a state with dominantly not more than one photon. We can calculate the value $p/q$ exactly for both states yielding
\begin{eqnarray}
\left.\frac{p}{q}\right|_\textrm{coh}&=&\frac{\langle\hat n\rangle}{\exp(\langle\hat n\rangle)-1-\langle\hat n\rangle},\\
\left.\frac{p}{q}\right|_\textrm{th}&=&\frac{1}{\langle\hat n\rangle}.
\end{eqnarray}
In comparison, the lower bounds based on equation~(\ref{eq.lower_bound_g2}) of the main text are
\begin{eqnarray}
\left.\frac{p}{q}\right|_\textrm{coh}&\geq&\frac{2\sqrt{2e^{-\langle\hat n\rangle}-1}}{1-\sqrt{2e^{-\langle\hat n\rangle}-1}},\quad \langle\hat n\rangle\leq \ln(2),\\
\left.\frac{p}{q}\right|_\textrm{th}&\geq&\frac{2\sqrt{1-\frac{4\langle\hat n\rangle}{1+\langle\hat n\rangle}}}{1-\sqrt{1-\frac{4\langle\hat n\rangle}{1+\langle\hat n\rangle}}},\quad \langle\hat n\rangle\leq 1/3.
\end{eqnarray}
Both cases are depicted in figure~\ref{fig.coh,th}. It can clearly be seen that for low excitation, these classical states also are detected using the effective second-order correlation function. Furthermore, our lower bounds appear as very good approximations for $\langle\hat n\rangle\lesssim 0.1$. Another way to look at this result is that when including the necessary information to calculate the SPP, it may not limit the fields to nonclassical states of light. 
\begin{figure}[h]
  \includegraphics[width=6cm]{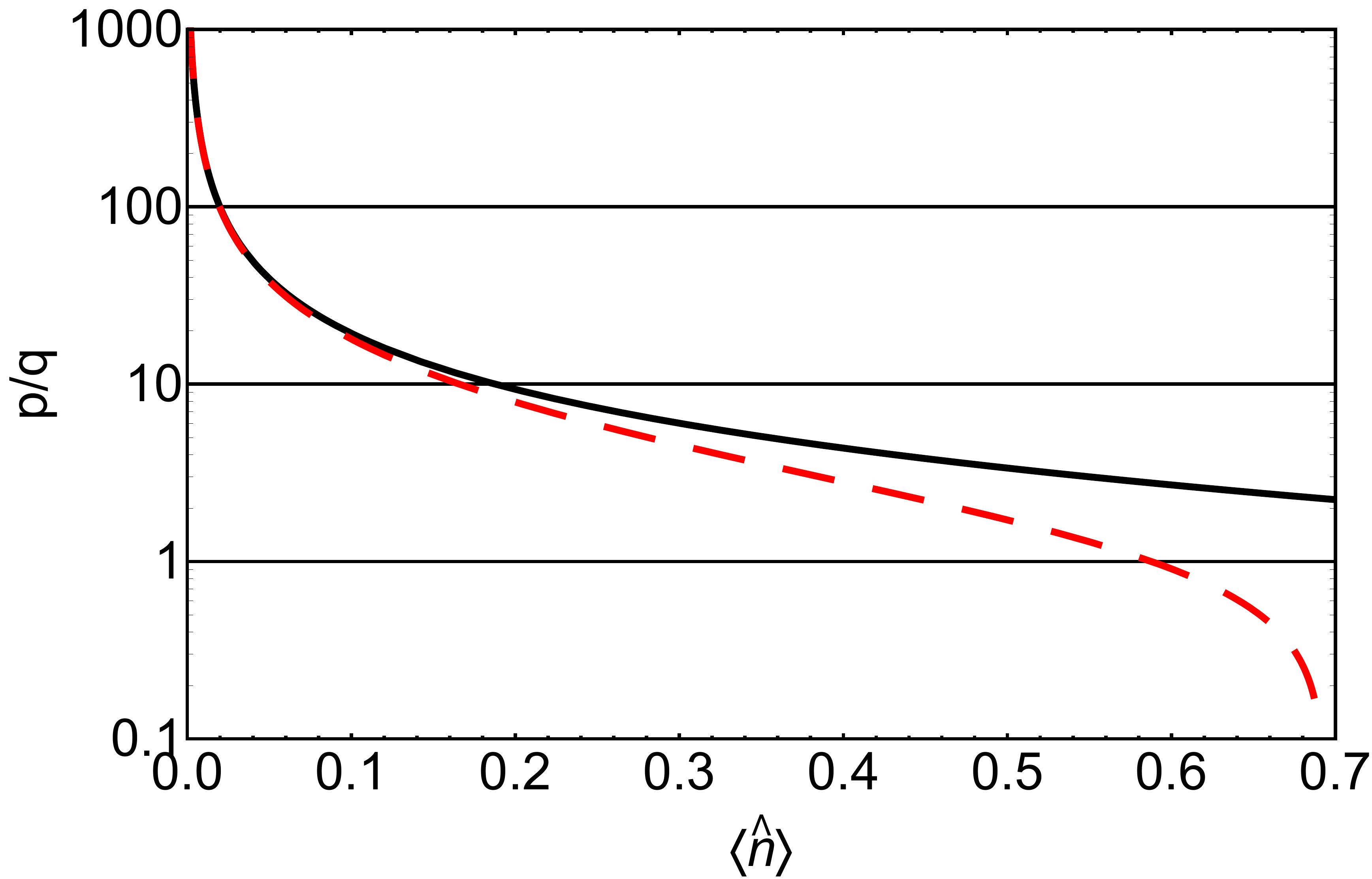}\hfill\includegraphics[width=6cm]{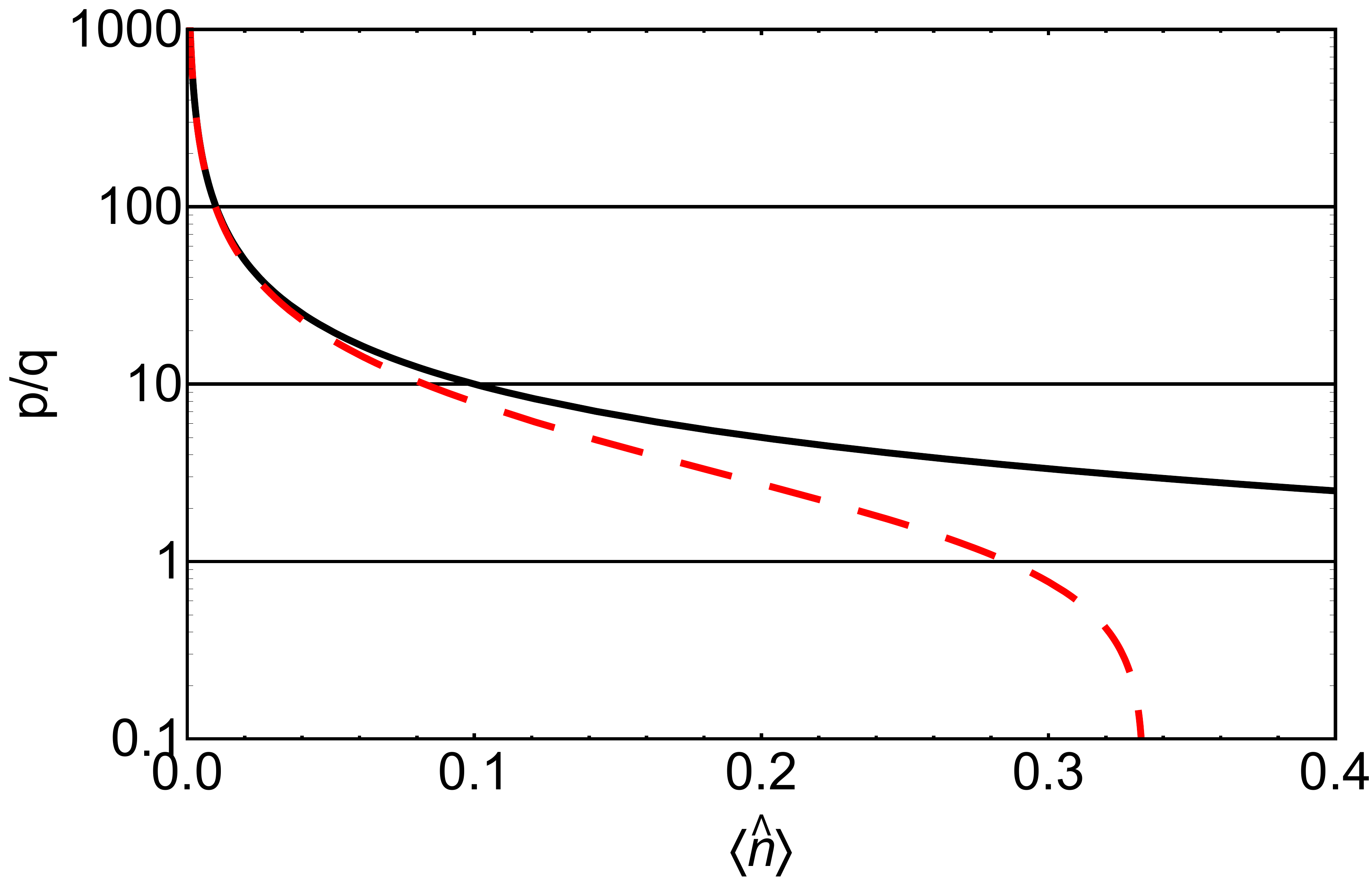}
  \caption{Exact formula (black solid) and lower bound (red dashed) of $p/q$ for a coherent state (left plot) and a thermal state (right plot).}\label{fig.coh,th}
\end{figure}

\section{Different interpretations based on $\tilde g^{(2)}$}\label{sec.alternative}
So far we have focused explicitely on the SMPP, as the major quantity to be gained from measuring $g^{(2)}$. In the following we give a few other interpretations of our results.
First, let us reconsider the SP purity $b=1-g^{(2)}$. Owing to the limitations we have shown above, a reasonable "purity" of SPP in the quantum state of light field can only be given in comparison to multi-photon projection. Thus, we may define a purity $\tilde b$ as the probability of obtaining SPs in a photon-number measurement, under the condition that more than zero photons appear at all. Within our notation $\tilde b$ then reads as
\begin{equation}
	\tilde b=\frac{p}{p+q}\geq\frac{C}{1+C}=\frac{2\sqrt{1-2\tilde g^{(2)}}}{1+\sqrt{1-2\tilde g^{(2)}}}\approx 1-\frac{\tilde g^{(2)}}{2}.
\end{equation}
We used $\tilde g^{(2)}\ll 1$ in the approximation. In case of no information on the vacuum contribution $x$ we again set $\tilde g^{(2)}=g^{(2)}$ and obtain a rather similar result as for the original SP purity $b$. Yet, the purity of SPs as defined by $\tilde b$ improves $b$ by a factor of two, even without knowledge of $x$. Note also that in case of no vacuum $x=0$, this SP purity is actually identical to a lower bound on $p$ itself.

Another way of interpreting this result is the upper bound on the multi-photon projection $q$. 
We easily see from rewriting equation~(\ref{eq.lower_bound_g2}) in terms of $q$ that
\begin{eqnarray}
		&\frac{1-q}{q}=&\frac{x+p}{q}\geq\frac{p}{q}\geq C,
\end{eqnarray}
from which an upper bound on $q$ and thus a lower bound on $1-q$ follows, as
\begin{eqnarray}
		1-q=x+p\geq\frac{C}{1+C}.\label{eq.newIP}
\end{eqnarray}
If for a quantum state of light $\hat\varrho$ one obtains $\tilde g^{(2)}(0)<1/2$, then there is a non-zero lower bound on the sum of projections on zero-photon and single-photon Fock state, 
\begin{equation}
	\langle0|\hat\varrho|0\rangle+\langle1|\hat\varrho|1\rangle\geq \frac{2\sqrt{1-2\tilde g^{(2)}}}{1+\sqrt{1-2\tilde g^{(2)}}}>0.
\end{equation}
The notion of very low multi-photon projection from low $g^{(2)}$-measurements has been an important aspect of research, see e.g.~\cite{Vuckevic2018,Yamamoto2001}.

At this point, it is worth comparing with the results from~\cite{Zubizarreta2017}. In that work it was shown that $g^{(2)}\leq1/2$ implies that the average photon number $\langle\hat n\rangle$ has to be smaller or equal to 2, the equality given for the Fock state $|2\rangle$. From this one can easily deduce upper bounds on the individual multi-photon projections $q_n=\langle n|\hat\varrho|n\rangle$, $n\geq2$, as
\begin{equation}
	n(n-1)q_n\leq\sum\limits_{n=2}^\infty n(n-1)q_n=\langle\hat n\rangle^2g^{(2)}\leq2,\quad \textrm{for } g^{(2)}\leq 1/2\label{eq.Zubibound}
\end{equation}
and thus $q_n\leq 2/[n(n-1)]$. Hence, a value of $g^{(2)}\leq 1/2$ implies $q_3\leq 1/3$, $q_4\leq1/6$ and so on.

In our derivation of equation~(\ref{eq.lower_bound_g2}), we used the monotonicity of $g^{(2)}$ with respect to increasing Fock states $|n\rangle$. One can conclude that for a fixed $g^{(2)}$ below the value for that Fock state, the highest possible projection $q_n$ is realized in a state containing only contributions from the single- and $n$-photon Fock state, the state given in equation~(\ref{eq.psilargeg2}). In that notation the quantity $q_n$ from above is just $1-p$ and we can calculate $q_3\leq0.134,q_4\leq0.057$ and so on.
For $n\gg1$ $q_n$ becomes
\begin{equation}
	q_n\leq \frac{g^{(2)}}{n(n-1)}\leq\frac{1}{2n(n-1)},\quad g^{(2)}<1/2.
\end{equation}
This is already a factor of 4 better than the upper boundary from~equation~(\ref{eq.Zubibound}). 
We can further improve the result by using the effective second-order correlation function $\tilde g^{(2)}$ instead of the bare $g^{(2)}$. For obtaining the maximal contribution of $q_n$ for $x\neq0$, one has to apply the corresponding conditions $n_2=n$ and $g_2=1-1/n$ in equation~(\ref{eq.p/qapp}). For example, given 50\% vacuum ($x=0.5$) and $g^{(2)}=1/2$ implies a maximal 3-photon projection $q_3=0.051$.

\section{Comparison with previous works}\label{sec.4}
This section is devoted to comparison with previous works from the solid-state community. While $x$ should be easily measurable in experiments without resorting to a full quantum-state reconstruction, it is so far usually not the focus of research. Thus, we will carefully extract a lower limit for $x$ from the data in the following references to not overestimate the effect of the vacuum. 

In the experiment performed in~\cite{Exp2009}, the authors use a quantum dot in a high-quality pillar micro cavity, obtaining for low cw-laser power and after subtracting experimental limitations $g^{(2)}\approx0.08$, which is already a very good value with $p/q\geq22$. However, as this is a weakly excited quantum dot, vacuum should be relevant. Taking the fit parameters of the experiment (Rabi frequency $\hbar\Omega=0.9$~$\mu$eV, lifetimes $T_2\approx2T_1=1150$~ps) and applying those to the same simple two-level model they used, we roughly find $x\approx0.58$, and thus $\tilde g^{(2)}\approx 0.034$ and $p/q\geq56$. In this case an already good SP source can be shown to be even better by evaluating the vacuum. Furthermore, as we have an explicite value for $x$, we can calculate the SPP $p$ from equation~(\ref{eq.abs_bounds}) as
\begin{equation}
0.41\leq p\leq 0.42.
\end{equation}
Thus, we get a very precise value for the real SP probability in a photon number measurement.

In comparison, consider~\cite{Santis16}, wherein the authors experimentally analyze a SP filter changing from SP characteristics ($g^{(2)}<1/2$) to coherent dynamics ($g^{(2)}=1$) by varying the input laser power. For low laser power $g^{(2)}\approx0.35$, which without knowledge of the vacuum only implies $p/q\geq2.5$. For this and the following example, we use the relation $x\geq 1-\langle\hat n\rangle$. Using figure~3a from that paper, we find for an input photon number of 0.1 that $\langle\hat n\rangle\leq0.1$ and thus $x\geq0.9$. This means, vacuum is the dominating contribution in this light field. Using the effective second-order correlation function, we can state $\tilde g^{(2)}\approx0.035$ and $p/q\geq54$. This is almost as good as the results of~\cite{Exp2009} above. Moreover, if we go to an input photon number of 0.9, they obtained $g^{(2)}\approx0.5$, but still $x\geq0.6$. Hence, $\tilde g^{(2)}\approx 0.2$, which is still a better SMPP ratio, than for the original low-power limit given by the authors. Our results show that not so good SP sources may only be disguised as such via strong vacuum contributions. 

\begin{table}[h]
\begin{tabular}{|l|c|c|c|c|c|c|}\hline
\textbf{Ref.} & $g^{(2)}$ & $p/q(x=0)$ & $x$ & $\tilde g^{(2)}$ & $p/q$ & $\tilde b$\\
\hline
\cite{Exp2009} & $0.08$ & 22 & $0.58$ & 0.034 & 56 & 98.2\%\\
\hline
\cite{Santis16}a & $0.35$ & 2.5 & $0.9$ & 0.035 & 54 & 98.2\%\\
\hline
\cite{Santis16}b & $0.5$ & 0 & $0.6$ & 0.2 & 6.87 & 87.3\%\\
\hline
\cite{Pagel15} & $4$ & 0 & $1-10^{-9}$ & $4\times 10^{-9}$ & $5\times10^8$ & 99.9999998\%\\
\hline
\end{tabular}
\caption{Comparison of the experimental and theoretical results of previous works with respect to SMPP ratio. The third column gives the lower bound on $p/q$ without knowledge of the vacuum contribution $(x=0)$, whereas the last two columns give that bound with knowledge of $x$ and the effective SP purity.}
\label{tab.1}
\end{table}

Now let us look at a theoretical example which does not aim at SPs at all~\cite{Pagel15}. Therein, the authors analyze a few emitters in a cavity and look at the output field. In particular, figures 4 and 6 of that work show the average photon number and $g^{(2)}$, respectively, for varying temperatures and cavity-emitter coupling strengths for the case of two emitters. Combining both figures and our above analysis we can estimate that a nonzero SPP is given in almost every point of the depicted state space, but most interesting is the upper left corner in part (a) of these figures, showing low temperature and high coupling strength. In this region $g^{(2)}\gtrsim 4$, which is far above the SP limit of 0.5. Yet, with an average photon number of the order of $10^{-9}$ it becomes virtually impossible to observe more than one photon at once, as $p/q\approx5\times10^8$. We have gathered all mentioned results in Tab.~\ref{tab.1}. They clearly show how much higher the quality of a SP source, at least with respect to the SMPP ratio may be than what the sole value of $g^{(2)}$ implies.

\begin{figure}
\includegraphics[width=6cm]{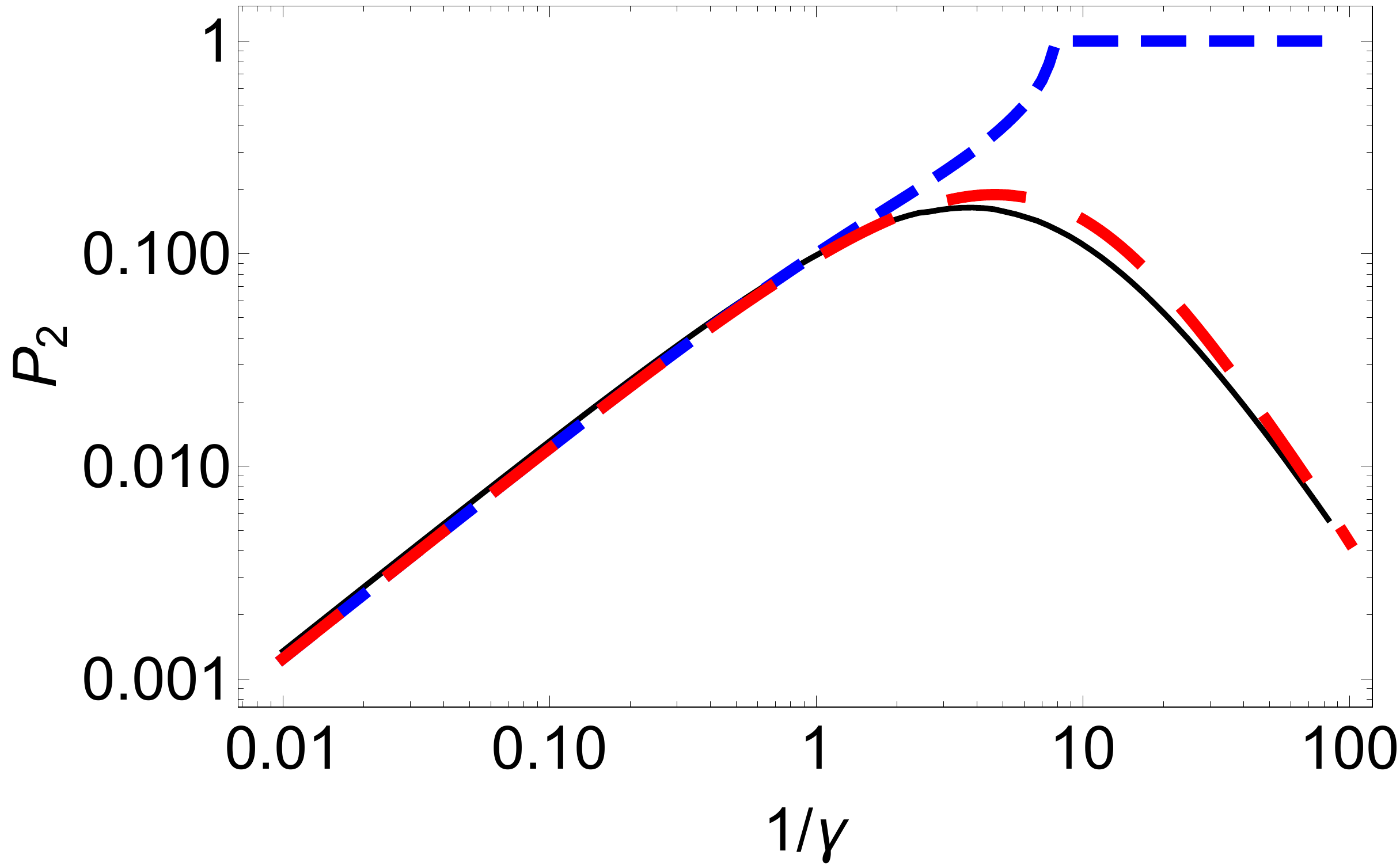}
\hfill
\includegraphics[width=6cm]{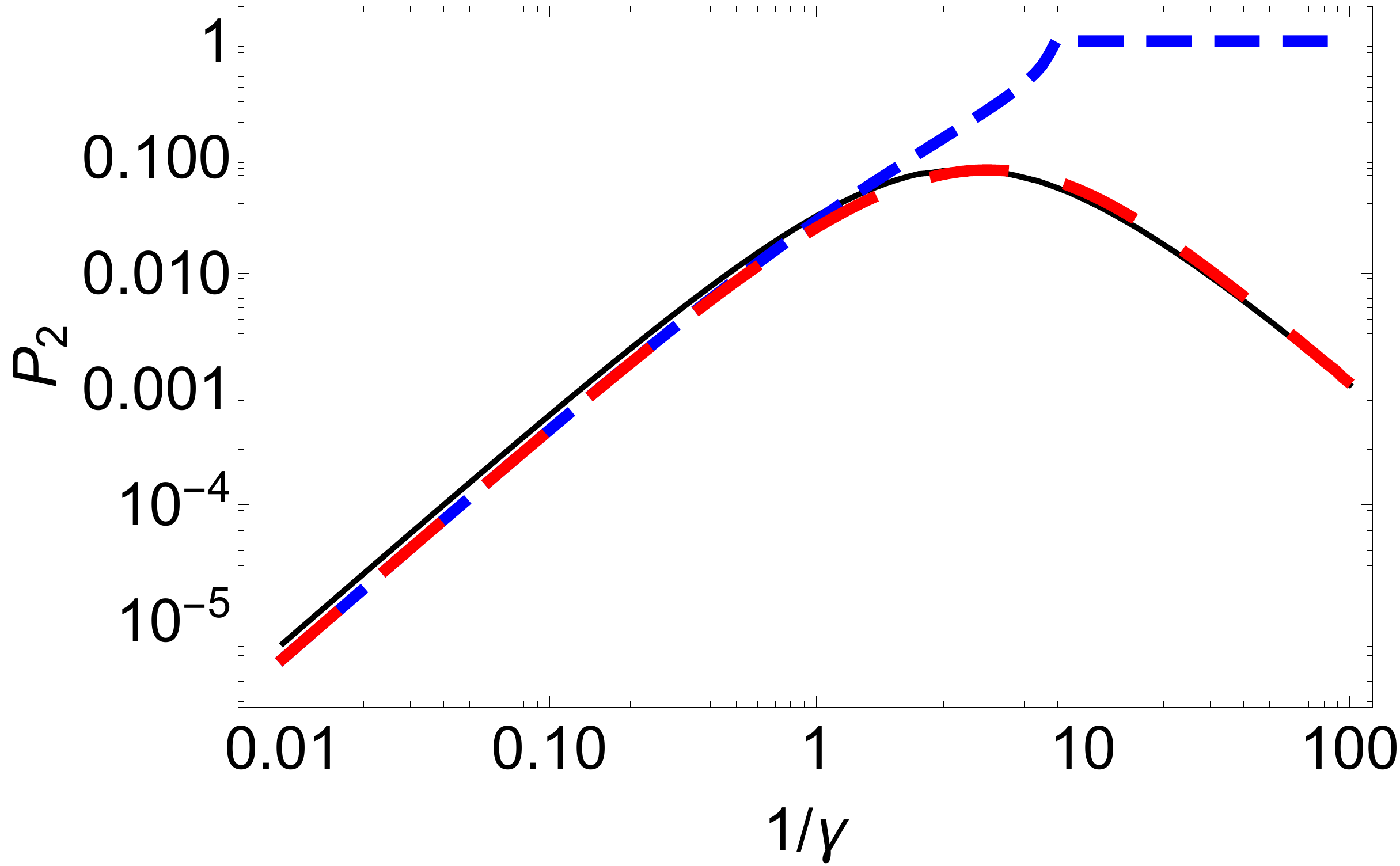}
\caption{Comparison of the results of~\cite{Vuckevic2018} with our description. Left: for the two-level system, Right: for the three-level system. The black solid line indicates the numerics for the two-photon contribution over increasing normalized excitation pulse length $1/\gamma$. Using the data of the authors for $g^{(2)}$ and $P_0=x$, we can compare with the upper bound for our $q\geq P_2$ for the case without (blue, narrow dashing) and with (red, wide dashing) taking into account vacuum.}\label{fig.Vuckevic}
\end{figure}

Finally, in a very recent theoretical study the authors compared two different quantum-dot SP setups and analyzed theoretically the two-photon contribution~\cite{Vuckevic2018}. Helpful for our comparison is that the authors present numerical results for $g^{(2)}$ as well as $x$, $p$, and the two-photon projection similar to our $q$, there denoted $P_0,P_1,$ and $P_2$, respectively. We consider figure~3 of ~\cite{Vuckevic2018}, wherin all these quantities for both setups are plotted as a function of laser-pulse length. Using equation~(\ref{eq.newIP}), we can determine an upper bound on $q$ and thus on $P_2$ to compare with the numerical simulation of the two-photon projection, see figure~\ref{fig.Vuckevic}. 

If information of the vacuum is not included, the result follows nicely the actual projection for short pulse lengths and correspondingly little vacuum. Around the maximum projection of $P_2$ at normalized pulse length of 3, vacuum becomes relevant and while the actual projection falls off, our upper bound grows up to one, when $g^{(2)}$ increases above 1/2. In contrast, including vacuum, for both setups the upper bound follows the actual Fock-state projection for all regions. In particular, for the two-level setup on the left, we see around the maximum some difference between the simulation and the upper bound, indicating an even higher Fock state projection than $n=2$. For the three-level setup, the upper bound stays close to the projection for all pulse lengths. Thus, in this setup there is virtually no higher Fock-state contribution. Finally, for completeness, it should be noted that for very short pulse lengths our upper bound falls slightly below the calculated projection. This deviation may come from numerical limitations or from the fact that for very short pulses the second-order correlation function is not based on steady-state correlations.

\section{Measurement of $\tilde g^{(2)}$}\label{sec.5}
Experimental limitations caused by strong vacuum contributions in weak light fields have been considered before. In~\cite{Hong2017} the authors discuss an optomechanical system with a dipole-like coupling between the optical photon and the mechanical phonon. Thus, a one-to-one relation between excited states of the two subsystems was established and single-phonon states were detected via single-photon measurement. The latter was perforemd by standard HBT experiment. Vacuum was a major issue in this work and hence, the authors circumvented this problem by employing post-selection technics, see figure~2 of~\cite{Hong2017}. In a first step, they detected the emission of a photon from the initialization process, before applying HBT measurements in the next step. Thus, the $g^{(2)}$ measurement was limited to the cases with no vacuum. 

From a theoretical point of view, this procedure is the inversion of the vacuum inclusion used in equation~(\ref{eq.vacincl}), removing the vacuum component $x$ from the original quantum state of light. Correspondingly, the determined $g^{(2)}$ of the post-selected state is equal to the effective second-order correlation function $\tilde g^{(2)}$ of the original quantum state of light. Thus, without explicite determination of the vacuum contribution $x$, we can obtain the effective second-order correlation function and thus calculate the lower limit for the SMPP ratio, equation~(\ref{eq.lower_bound_g2}).

However, there are two drawbacks to this method. On the one hand, as $x$ itself is not determined, we still do not obtain bounds for the SPP $p$, rather a lower bound for $\tilde p=p/(1-x)$. On the other hand, we do not analyze the original fields, but only the state without vacuum. While a nonzero SPP can easily be detected in this way, the anti-bunching character found from $g^{(2)}<1$ may be completely lost, unless we know that the original $g^{(2)}$ fullfilled this condition, see the examples in section~\ref{sec.kS}. Hence, we conclude that the values with- and without vacuum removal must both be determined, or, in other words, the vacuum contribution $x$ is a necessary ingredient to estimate the actual SPP in the state.

\section{Conclusions and Outlook}\label{sec.out}
We have analyzed the gain of information from $g^{(2)}$ with respect to SPs. Any quantum state, for which the second-order correlation function falls below $1/2$, has a nonzero projection on the SP Fock state. The amplitude $p$ of this projection is arbitrary, independent of $g^{(2)}(0)$. However, one can extract a lower bound on the SMPP ratio. A vacuum contribution in the quantum state of light artficially increases the value of $g^{(2)}$, cloaking actual SPP. Thus, we proposed an effective second-order correlation function $\tilde g^{(2)}$, which takes the influence of vacuum into account and also yields lower and upper bounds on $p$, when combined with the information about the vacuum projection. We considered the SP purity as a standard figure-of-merit in experiments and reinterpreted it within our results. Comparison with other experimental and theoretical results indicates that there are many more SP light sources, where the SMPP ratio is much higher than expected, due to the vacuum contributions. We also provided a measurment scheme for $\tilde g^{(2)}$, which, however, may yield artificial nonclassicality of the quantum state of light.

These vacuum contributions entered our derivation quite naturally, and its physical origin was only explained afterwards. As indicated by the chosen examples the average photon number is an often determined quantity that may also provide further results on the SP character, compare also~\cite{Zubizarreta2017}. 
Other figures of merit for SP sources include indistinguishability~\cite{Santori2002} and coalescence~\cite{Polyakov2011} of different SPs, which require more emphasis on the decohering processes and thus the Hamiltonian and dissipative structure of the system at hand.
Each of these quantities yield their own additional information, but require an individual derivation akin in size to the case discussed in this work. Performing and tracking these derivations is intended as future work.

\section*{Acknowledgments}
The author gives special thanks to K. M\"uller for providing the raw data used to generate figure~\ref{fig.Vuckevic}. 
This work was in part supported by the DFG through collaborative research center 652, "Strong correlations and collective effects in radiation fields", by the EU through the H2020-FETOPEN grant No. 800942 640378 (ErBeStA), and by the
DNRF through a Niels Bohr Professorship to Thomas Pohl.

\section*{References}
%\bibliography{bib_g2}

\providecommand{\newblock}{}

\end{document}